# Implications of Inter-Rater Agreement on a Student Information Retrieval Evaluation

**Philipp Schaer, Philipp Mayr, Peter Mutschke**
GESIS – Leibniz Institute for the Social Sciences
53113, Bonn, Germany
philipp.schaer@gesis.org

## Abstract

This paper is about an information retrieval evaluation on three different retrieval-supporting services. All three services were designed to compensate typical problems that arise in metadata-driven Digital Libraries, which are not adequately handled by a simple *tf-idf* based retrieval. The services are: (1) a co-word analysis based query expansion mechanism and re-ranking via (2) Bradfordizing and (3) author centrality. The services are evaluated with relevance assessments conducted by 73 information science students. Since the students are neither information professionals nor domain experts the question of inter-rater agreement is taken into consideration. Two important implications emerge: (1) the inter-rater agreement rates were mainly fair to moderate and (2) after a data-cleaning step which erased the assessments with poor agreement rates the evaluation data shows that the three retrieval services returned disjoint but still relevant result sets.

## 1. Introduction

Metadata-driven Digital Libraries (DL) face three typical difficulties: (1) the vagueness between search and indexing terms, (2) the information overload by the amount of result records returned by information retrieval systems, and (3) the problem that pure term frequency based rankings, such as term frequency – inverse document frequency (*tf-idf*), provide results that often do not meet user needs [Mayr *et al.*, 2008]. To overcome these limitations the DFG-funded project "Value-Added Services for Information Retrieval" (IRM[1]) is doing research on an overall approach to use computational science models as enhanced search stratagems [Bates, 1990] within a scholarly retrieval environment, which might be implemented within scholarly information portals.

To show that a user's search improves by using these model-driven search services when interacting with a scientific information system, an information retrieval evaluation was conducted. Since the assessors in this experiment were neither information professionals nor domain experts we had especially looked at the level of agreement between the different assessors. This also is our special interest since in big evaluation campaigns like CLEF only a minority of the evaluations made their evaluations with regard to the inter-rater agreement. In the CLEF campaign of 2009 no analysis of inter-rater agreement was made in the classical ad-hoc track [Ferro and Carol, 2009] and only one to four assessors had to judge the pooled documents. The only track that made these inter-rater analyses was the medical imaging track [Müller *et al.*, 2009], but they only had two assessors per topic. These numbers are quite low compared to our evaluation where for some topics there were up to 15 assessors. With the high number and the non-professional background of our assessors a statistical measure is needed to rate the reliability and consistency of agreement between the different assessments.

Several metrics have been developed to measure inter-grader reliability, like mean overlap or Fleiss' Kappa, which will be briefly presented in the next section. After that we will introduce the three services that are based on the principles of co-words, core journals and centrality of authors. The basic assumptions and concepts are presented. The conducted evaluation and study with 73 participants is described in the following section. The paper closes with a discussion of the observed results.

## 2. Inter-rater Agreement

The reliability and consistency of agreement between different assessments can be measure by a range of statistical metrics. Two of these metrics to measure inter-grader reliability are for example mean overlap or Fleiss' Kappa.

### 2.1. Mean Overlap and Overall Agreement

In the early TREC conferences (namely till TREC 4) a rather simple method was used to measure the amount of agreement among different assessors by calculating the overlap between the different assessors judgements. Overlap was defined by the size of the intersection divided by the size of the union of the relevant document sets. Despite the relatively high average overlap, Voorhees [2000] reported that there were significant different judgements for some topics. She reported that in TREC topic 219 for example there was no single intersection between two assessors at all. Therefore the mean overlap in this TREC topic was between 0.421 and 0.494 comparing two assessors directly and 0.301 comparing the set of all assessors together.

This measure was discarded after TREC 4. One of the reasons was that this measure gets very unstable when more than three assessors are taken into account.

---

[1] http://www.gesis.org/irm

## 2.2. Fleiss' Kappa

Fleiss' Kappa is a measure of inter-grader reliability – based on Cohen's Kappa – for nominal or binary ratings [Fleiss, 1971]. The Kappa value can be interpreted as the extent to which the observed amount of agreement among raters exceeds what would be expected if all raters made their ratings completely randomly.

Fleiss' Kappa is given in equations 1 to 3:

$$\kappa = \frac{\overline{P} - \overline{P}_e}{1 - \overline{P}_e} \quad (1)$$

where

$$\overline{P} = \frac{1}{Nn(n-1)} \sum_{i=1}^{N} \sum_{j=1}^{k} n_{ij}(n_{ij} - 1) \quad (2)$$

$$\overline{P}_e = \sum_{j=1}^{k} \left( \frac{1}{Nn} \sum_{i=1}^{N} n_{ij} \right)^2 \quad (3)$$

$N$ is the total number of subjects (e.g. documents to be assessed); $n$ is the number of judgments per subject (raters); $k$ is the number of response categories.

Kappa scores can range from 0 (no agreement) to 1.0 (full agreement). Landis and Koch [1977] suggest interpreting the score as followed: $\kappa<0$ = poor agreement, $0\leq\kappa<0.2$ = slight agreement, $0.2\leq\kappa<0.4$ = fair agreement, $0.4\leq\kappa<0.6$ = moderate agreement, $0.6\leq\kappa<0.8$ = substantial agreement, $0.8\leq\kappa\leq1$ = (almost) perfect agreement. These interpretations are not generally accepted and other interpretations are possible. Greve and Wentura [1997] suggest to interpret scores $\kappa<0.4$ as "not be taken too seriously" and values of $0.4\leq\kappa<0.6$ as acceptable. $0.75\leq \kappa$ seems good up to excellent.

## 3. Evaluated System and Services

All proposed models are implemented in a live information system using (1) the Solr search engine, (2) Grails Web framework and (3) Recommind Mindserver to demonstrate the general feasibility of the approaches. Solr is an open source search platform from the Apache Lucene project[2], which uses a *tf-idf*-based ranking mechanism[3]. The Mindserver is a commercial text categorization tool, which was used to generate the query expansion terms. Both Bradfordizing and author centrality as re-rank mechanism are implemented as plugins to the open source web framework Grails, which is the glue to combine the different modules and to offer an interactive web-based prototype[4].

### 3.1. Query Expansion by Search Term Recommendation

When using search in an information system a user has to come up with the "correct" terms to formulate his query. These terms have to match the terms used in the documents or in the description of the documents to get an appropriate result. Especially in the domain of metadata-driven Digital Libraries this is long known as the language problem in IR [Petras 2006].

The Search Term Recommender (STR) is based on statistical co-word analysis and builds associations between free terms (i.e. from title or abstract) and controlled terms (i.e. from a thesaurus). Controlled terms are assigned to the document during a professional indexation and enrich the available metadata on the document. The co-word analysis implies a semantic association between the free and the controlled terms. The more often terms co-occur in the text the more likely it is that they share a semantic relation. The commercial classification software Recommind Mindserver, which is based on Support Vector Machines (SVM) and Probabilistic Latent Semantic Analysis (PLSA), calculated these associations. The software was trained with the SOLIS database to best match the specialized vocabularies used in the Social Sciences.

When used as an automatic query expansion service the Mindserver's top n=4 suggested terms were taken to enhance the query. The query was expanded by simply OR-ing them. If we take a sample query on Poverty in Germany, it is expanded to:

```
povert* AND german* →

(povert* AND german*) OR "poverty" OR
"Federal Republic of Germany" OR "so-
cial assistance" OR "immiseration"
```

### 3.2. Re-Ranking by Bradfordizing

The Bradfordizing re-ranking service addresses the problem of oversized result sets by using a bibliometric method. Bradfordizing re-ranks a result set of journal articles according to the frequency of journals in the result set such that articles of core journals – journals which publish frequently on a topic – are ranked higher than articles from other journals. This way the central publication sources for any query are sorted to the top positions of the result set [Mayr, 2009].

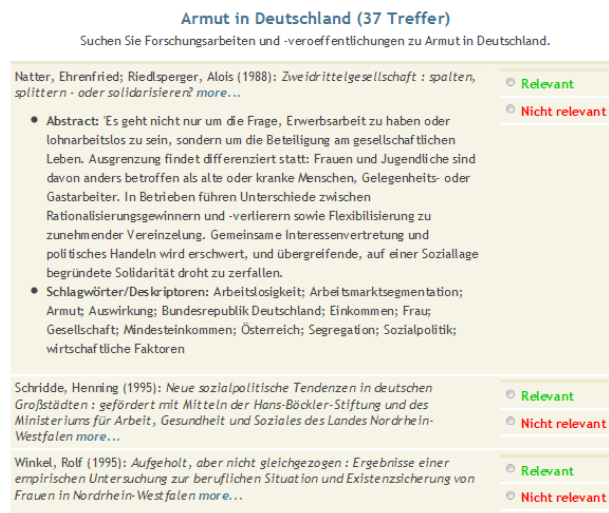

Figure 1: Screenshot of the web-based assessment tool. The users were shown the name and description of the task (top) and a number of documents to assess. The documents (middle) had author, publication year, title, abstract and keywords, which the assessors could use to judge the documents relevant or non-relevant (right).

---
[2] http://lucene.apache.org/Solr/
[3] http://lucene.apache.org/java/2_4_0/scoring.html
[4] http://www.gesis.org/beta/prototypen/irm/

|       |                               |                                                                                                                          | Fleiss' Kappa |    |   |       | Mean overlap |       |
|-------|-------------------------------|--------------------------------------------------------------------------------------------------------------------------|---------------|----|---|-------|--------------|-------|
| topic | title                         | description                                                                                                              | n             | N  | k | κ     | >=0.8        | =1    |
| 83    | Media and War                 | Find documents on the commentatorship of the press and other media from war regions.                                     | 15            | 40 | 2 | 0.522 | 0.727        | 0.225 |
| 84    | New Media in Education        | Find documents reporting on benefits and risks of using new technology such as computers or the Internet in schools.     | 11            | 40 | 2 | 0.304 | 0.5          | 0.15  |
| 88    | Sports in Nazi Germany        | Find documents about the role of sports in the German Third Reich.                                                       | 6             | 40 | 2 | 0.528 | 0.75         | 0.425 |
| 93    | Burnout Syndrome              | Find documents reporting on the burnout syndrome.                                                                        | 10            | 40 | 2 | 0.411 | 0.8          | 0.35  |
| 96    | Costs of Vocational Education | Find documents reporting on the costs and benefits of vocational education.                                              | 2             | 40 | 2 | 0.488 | 0.775        | 0.775 |
| 105   | Graduates and Labour Market   | Find documents reporting on the job market for university graduates.                                                     | 5             | 40 | 2 | 0.466 | 0.675        | 0.525 |
| 110   | Suicide of Young People       | Find documents investigating suicides in teenagers and young adults.                                                     | 5             | 40 | 2 | 0.222 | 0.625        | 0.425 |
| 153   | Childlessness in Germany      | Information on the factors for childlessness in Germany                                                                  | 10            | 40 | 2 | 0.202 | 0.325        | 0.175 |
| 166   | Poverty in Germany            | Research papers and publications on poverty and homelessness in Germany.                                                 | 9             | 40 | 2 | 0.438 | 0.5          | 0.25  |
| 173   | Propensity towards violence among youths | Find reports, cases, empirical studies and analyses on the capacity of adolescents for violence.              | 10            | 40 | 2 | 0.411 | 0.55         | 0.2   |
|       |                               |                                                                                                                          |               |    | avg. | *0.4* | *0.622*  | *0.35* |

Table 1: Ten CLEF topics and the corresponding inter-grader reliability expressed by Fleiss' Kappa and mean overlap per topic. The mean overlap is calculated using two different thresholds (>=0.8 means that an intersection rate of 80% is counted and =1 means that only perfect matches are counted).

In a first step the search results are filtered with their ISSN, since ISSNs are proper identifiers for journals. The next step aggregates all results with the same ISSN. For this step the build-in faceting mechanism of Solr is used. The journal with the highest ISSN facet count gets the top position in the results; the second journal gets the next position, and so on. In the last step, each documents rank (given through Solr's internal ranking) is boosted by the frequency counts of the journals. This way all documents from the journal with the highest ISSN facet count are sorted to the top.

### 3.3. Re-Ranking by Author Centrality

Author centrality is another way of re-ranking result sets. Here the concept of centrality of authors in a network is an approach for the problem of large and unstructured result sets. The intention behind this ranking model is to make use of knowledge about the interaction and cooperation behavior in fields of research. The model assumes that the relevance of a publication increases with the centrality of their authors in co-authorship networks. The user is provided with publications of central authors when the result set obtained is re-ranked by author centrality.

The re-ranking service calculates a co-authorship network based on the result set to a specific query. Centrality of each single author in this network is calculated by applying the betweenness measure and the documents in the result set are ranked according to the betweenness of their authors so that publications with very central authors are ranked higher in the result list. Since the model is based on a network analytical view it differs greatly from text-oriented ranking methods like *tf-idf* [Mutschke, 2004].

## 4. Evaluation

We conducted a user assessment with 73 information science students who used the SOLIS database[5] with 369,397 single documents on Social Science topics to evaluate the performance of the three presented services. The documents include title, abstract, controlled keywords etc. The assessment system, which was built on top of the IRM prototype described earlier and all documents were in German. All written examples in this paper are translated.

### 4.1. Method

A standard approach to evaluate Information Retrieval systems is to do relevance assessments, where – in respect to a defined information need – documents are marked as relevant or not relevant. There are small test collections (like Cranfield) where all containing documents are judged relevant or not relevant. For large collections (like TREC, CLEF etc.) this is not possible, so only subsets are assessed. Only the top n documents returned by the different retrieval systems are assessed. Pooling is used to disguise the origin of the document [Voorhees and Harman, 2005].

In our assessment the participants were given a concrete search task, which was taken from the CLEF corpus. After a briefing each student had to choose one out of ten different predefined topics (namely CLEF topics 83, 84, 88, 93, 96, 105, 110, 153, 166 and 173). Topic title and the description were presented to form the information need (cp. table 1 and figure 1).

---
[5] SOLIS is accessable via the Sowiport portal http://www.gesis.org/sowiport

The pool was formed out of the top n=10 ranked documents from each service and the initial *tf-idf* ranked result set respectively. Duplicates were removed, so that the size of the sample pools was between 34 and 39 documents each. The assessors could choose to judge relevant or not relevant (binary decision) – in case they didn't assess a document this document was ignored in later calculations.

### 4.2. Data Set

The 73 assessors, who were information science students, did 43.78 assessments in average. They did 3,196 single relevance judgments in total. Only 5 participants didn't fill out the assessment form completely. Since every assessor could freely choose from the topics the assessments are not distributed evenly. Topic 83 was picked 15 times – topic 96 twice. It was a conscious decision to allow each assessor to freely choose a topic he or she is familiar with, but this had direct consequences on the number of single assessments (cp. table 2).

## 5. Results

Since the assessors in this experiment were neither information professionals nor domain experts the results should be discussed with a special interest on the level of agreement between the assessors. To rate the reliability and consistency of agreement between the different assessments we used mean overlap and Fleiss' Kappa, described in section 2. Besides that precision of each topic/service combination and the intersection of the different result sets was calculated.

### 5.1. Mean Overlap and Overall Agreement

Normally only a true overlap is counted. When comparing the judgements of two assessors they can either agree or disagree. When dealing with more than two assessors the situation gets difficult: What to do if 14 assessors agree but one disagrees? To take majority decision into account we applied two different thresholds calculating the mean overlap: 1 and 0.8. Only perfect matches count when a threshold of 1 is applied: This means that all assessors have to agree 100%. Here the measured mean overlap (140 intersections where all assessors agreed 100% in a total union set of 400) was 0.35 (cp. table 1). This value can be compared to the reported values from the selected TREC topic (see section 2.1) where the overlap was between 0.421 and 0.494 comparing two assessors directly and 0.301 comparing the set of all three assessors together.

The more assessors per topic the lower the mean overlap gets. This is quite natural since 100% agreement is easier to obtain between 2 assessors (for example for topic 96) than for 15 assessors (like for example in topic 83). Therefore a second threshold of 0.8 was applied. Here all judgements with only slight disagreement where also taken into consideration. This had direct implications on the average overlap value over all assessments. The average overlap rate was 0.622.

### 5.2. Fleiss' Kappa

The equation for Fleiss' Kappa is given in section 2.2. In our study N was always forty (ten documents for each value added service – here assessments for duplicates count for all services that returned it) and k was always two (binary decision).

All Kappa scores in our experiment range between 0.20 and 0.52, which are fair up to moderate levels of agreement or mainly acceptable in the more conservative interpretation (cp. table 1). The average Kappa value was 0.4.

### 5.3. Precision

The precision *P* was calculated by equation 4:

$$P = \frac{|\{rel\} \cap \{ret\}|}{|\{ret\}|} \quad (4)$$

*P* was calculated for each topic and service, where |{rel}| is the number of all relevant assessed documents and |{ret}| is the number of all retrieved documents which were in the pool (relevant and not relevant). All unfiltered precision values and numbers of relevance assessments can be seen in table 2 and figure 2.

The average precision of the STR was highest (68%) compared to the baseline from the SOLR system (57%). The two alternative ranking methods Bradfordizing (BRAD) and author centrality (AUTH) scored 57% and 60% respectively.

We calculated additional average precision values and left out the topics marked unstable by the mean overlap values (those with a value of <=0.35 in case of a 100% match). Here the topics 83, 84, 153, 166 and 173 were left out. This was not considered useful since only half the data set remains after this data clearance step. Nevertheless calculating the average precision for each service without the five topics would have been: AUTH: 63%, BRAD: 63%, SOLR 61% and STR: 65%.

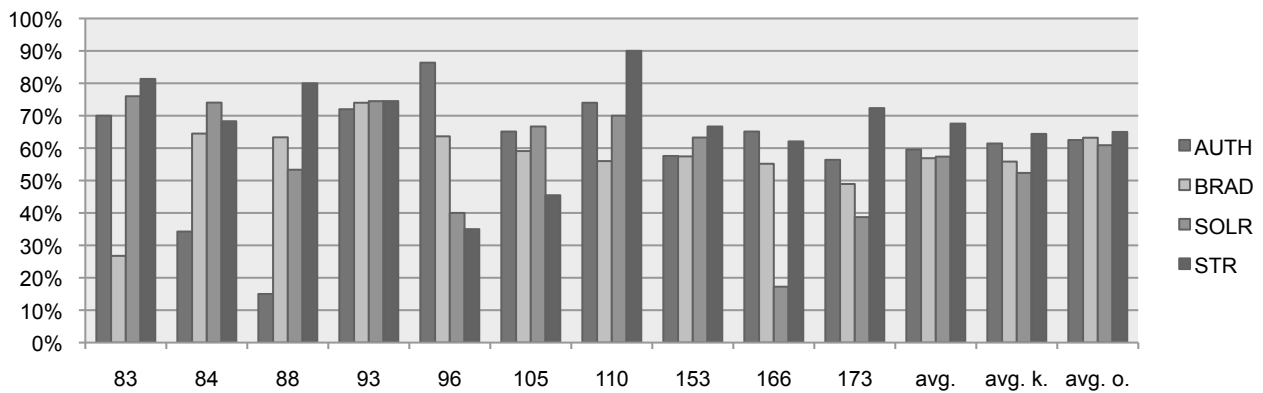

Figure 2: Precision for each topic and service (Relevance assessments per topic / total amount of single assessments). The three average precision values are (1) avg.: unfiltered values, (2) avg. k.: average computed with a kappa threshold of 0.4 and (3) avg. o.: average computed with a overlap threshold of 0.35.

Considering the Kappa values the relevance judgments from topics 84, 110 and 153 were below the threshold of 0.40 and so they might have been dropped. Calculating the average precision for each service without the three unstable topics would have been: AUTH: 61%, BRAD: 56%, SOLR 52% and STR: 64% (cp. table 2).

## 5.4. Intersection of result sets

A comparison of the intersection of the relevant top 10 document result sets between each pair of retrieval service shows that the result sets are nearly disjoint. 400 documents (4 services * 10 per service * 10 topics) only had 36 intersections in total (comp. figure 3). Thus, there is no or very little overlap between the sets of relevant top documents obtained from different rankings.

AUTH and SOLR as well as AUTH and BRAD have just three relevant documents in common (for all 10 topics), and AUTH and STR have only five documents in common. BRAD and SOLR have six, and BRAD and STR have five relevant documents in common. The largest, but still low overlap is between SOLR and STR, which have 14 common documents.

## 6. Discussion

The discussion will focus on two different aspects: (1) the results of the inter-rater agreement and our implications to our evaluation setup and (2) the evaluation itself and the outcomes regarding precision and intersection of result sets.

### 6.1. Inter-rater agreement

The central question we have to consider is to what amount the evaluation is significant before we can interpret the results in form of precision values. At first our results looked promising, when comparing them to the mean overlap agreement rate from the TREC studies, especially when taking the high number of assessors per topics into account. After we had a look at the mean over-lap and Kappa values we had to differentiate more on the single topics since their performance was not simply comparable.

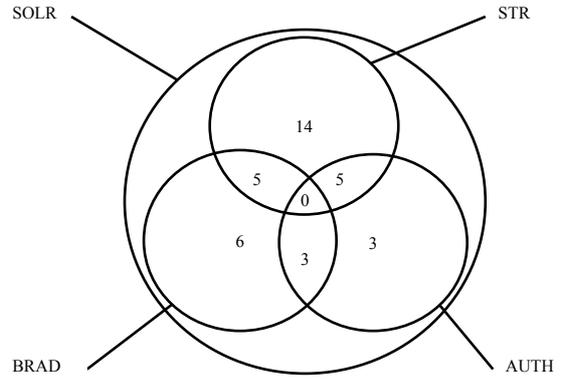

Figure 3: Intersection of top n=10 documents from all topics and services (total of 400 documents). The number 3 in the lower right circle for example means that only 3 documents in the result set returned by the SOLR system and the result set returned by the author centrality service AUTH are the same.

Taking the non-professional assessors and the Kappa values in consideration we have to ask ourselves how reliable our evaluations are. The general setup is not unusual: Other studies also relied on non-professional assessors and showed promising results: Al-Maskari *et al.* [2008] observed a rather high overall agreement between official TREC and non-TREC assessors in a study with 56 participants and 4399 single document assessments in total. The agreement changed due to the different topics and the actual ranking position of the assessed document and was between 75 and 82% if both relevant and irrelevant documents assessments were counted. Alonso and Mizzaro [2009] compared official TREC assessments with anonymous assessors from Amazon Mechanical Turk. They observed a rather small sample of 10 participants and 290 judgments. For the relevant documents the average across of participants was 91% and in case of not relevant documents the average was 49%.

|  | non relevant | | | | relevant | | | | precision | | | |
|---|---|---|---|---|---|---|---|---|---|---|---|---|
| topic | AUTH | BRAD | SOLR | STR | AUTH | BRAD | SOLR | STR | AUTH | BRAD | SOLR | STR |
| 83 | 42 | 104 | 36 | 25 | 98 | 38 | 114 | 109 | 0.70 | 0.27 | 0.76 | 0.81 |
| 84 | 71 | 38 | 27 | 26 | 37 | 69 | 77 | 56 | 0.34 | 0.64 | 0.74 | 0.68 |
| 88 | 51 | 22 | 28 | 12 | 9 | 38 | 32 | 48 | 0.15 | 0.63 | 0.53 | 0.80 |
| 93 | 28 | 26 | 25 | 26 | 72 | 74 | 73 | 76 | 0.72 | 0.74 | 0.74 | 0.75 |
| 96 | 3 | 8 | 12 | 13 | 19 | 14 | 8 | 7 | 0.86 | 0.64 | 0.40 | 0.35 |
| 105 | 15 | 18 | 15 | 24 | 28 | 26 | 30 | 20 | 0.65 | 0.59 | 0.67 | 0.45 |
| 110 | 13 | 22 | 15 | 5 | 37 | 28 | 35 | 45 | 0.74 | 0.56 | 0.70 | 0.90 |
| 153 | 42 | 40 | 36 | 32 | 57 | 54 | 62 | 64 | 0.58 | 0.57 | 0.63 | 0.67 |
| 166 | 30 | 39 | 72 | 33 | 56 | 48 | 15 | 54 | 0.65 | 0.55 | 0.17 | 0.62 |
| 173 | 41 | 48 | 57 | 26 | 53 | 46 | 36 | 68 | 0.56 | 0.49 | 0.39 | 0.72 |
| avg. | | | | | | | | | *0.60* | *0.57* | *0.57* | *0.68* |
| avg.(kappa >= 0.4) | | | | | | | | | *0.61* | *0.56* | *0.52* | *0.64* |
| avg. (overlap >= 0.35) | | | | | | | | | *0.63* | *0.63* | *0.61* | *0.65* |

Table 2: Relevance judgments for each topic and service (total number of non-relevant and relevant judgments) and the calculated precision values (with different thresholds applied).

Nevertheless a mandatory step should be to sort out subsets of the assessment data where Kappa values (or other reliable measures) are below a certain threshold. Unfortunately this is not done in all studies.

## 6.2. Precision and Intersection

Comparing the precision values of the different approaches the three services in average all performed better (in case of the filtered topic list) or at least same (when counting all assessments) as the naïve *tf-idf* baseline. This is true for all average precision values calculated regarding all kinds of thresholds and data clearing.

When inspected on a per-topic basis the performance is more diverse. While the STR outperforms in nearly all cases the other services had topics where their precision was significantly smaller compared to the other sources. For topic 83 the Bradfordizing re-ranking service couldn't sort the result set appropriately while the author centrality service couldn't adequately handle topic 88.

Discussing the results of the two proposed re-ranking methods Bradfordizing and author centrality brings up two central insights: (1) users get new result cutouts with other relevant documents which are not listed in the first section (first n=10 documents) of the original list and (2) Bradfordizing and author centrality can be a helpful information service to positively influence the search process, especially for searchers who are new on a research topic and don't know the main publication sources or the central actors in a research field. The STR showed an expected behavior: While the result set in total increases the first n=10 hits are more precise. This is quite normal in query expansion scenarios where the high descriptive power of the controlled terms that are added to the query increases the precision [Efthimiadis, 1996]. This is an indicator for the high quality of the semantic mapping between the language of scientific discourse (free text in title and abstract) and the language of documentation (controlled thesauri terms).

The very low overlap of the result sets as described in section 5.4 confirms that the value-added services proposed provide a quite different view to the document space: Not only from a term-centric view proposed by *tf-idf* (with or without query expansion mechanism) but also from a more person- or journal-centric perspective. Which is even more interesting since the average precision values didn't differ as much as one might have expected.

## 7. Outlook

After the evaluation and the analysis regarding inter-rater agreement two important implications emerge: (1) the inter-rater agreement rates were mainly fair to moderate and therefore showed a general feasibility of a non-experts evaluation and (2) after a data-cleaning step which erased the assessments with very poor agreement rates the evaluation data showed that the three retrieval services returned disjoint but still relevant result sets. The services provide a particular view to the information space that is quite different from traditional retrieval methods.

Although the results looked promising our next step is a new relevance assessment with scientist and domain experts to have a direct comparison and to reinsure our observations.

## Acknowledgments

Vivien Petras and Philipp Mayr supervised the 73 students in their courses at Humboldt University and University of Applied Science Darmstadt. Hasan Bas implemented the assessment tool during his internship with GESIS.

This work was funded by DFG (grant number INST 658/6-1).